\documentclass[%
reprint,
showpacs,
amsmath,amssymb,
aps,
prl,
floatfix,
]{revtex4-1}

\usepackage{dcolumn}

\usepackage[ansinew]{inputenc}
\usepackage[english]{babel}%
\usepackage{graphicx}
\usepackage{textcomp}

\usepackage{bm}
\usepackage{xcolor}
\usepackage[]{units}
\usepackage{changes}
\usepackage{soul}
\usepackage{xcolor}
\usepackage{hyperref}
\addto\captionsenglish{}

\begin{document}

\title{Rotational cooling of trapped polyatomic molecules}

\author{Rosa Gl\"ockner}
\author{Alexander Prehn}%
\author{Barbara G. U. Englert}%
\author{Gerhard Rempe}%
\author{Martin Zeppenfeld}%
 \email{Martin.Zeppenfeld@mpq.mpg.de}
\affiliation{Max-Planck-Institut f\"ur Quantenoptik, Hans-Kopfermann-Str.1, D-85748 Garching, Germany}%

\begin{abstract}{
Controlling the internal degrees of freedom is a key challenge for applications of cold and ultracold molecules. Here, we demonstrate rotational-state cooling of trapped methyl fluoride molecules (CH$_3$F) by optically pumping the population of 16 $M$-sublevels in the rotational states ${J{=}3,4,5 \textrm{, and }6}$ into a single level. By combining  rotational-state cooling with motional cooling, we increase the relative number of molecules in the state $J{=}4$, $K{=}3$, $M{=}4$ from a few percent to over 70\%, thereby generating a translationally cold ($\approx \unit[30]{mK}$) and nearly pure state ensemble of about 10$^6$ molecules. Our scheme is extendable to larger sets of initial states, other final states and a variety of molecule species, thus paving the way for internal-state control of ever larger molecules.}
\end{abstract}

\maketitle

Motivated by a multitude of applications ranging from quantum chemistry to many-body physics~\cite{ANDRE2006,Micheli2006,Hudson2011,Lemeshko2013a}, recent years have witnessed an immense effort to generate cold and ultracold ensembles of polar molecules~\cite{Ni2008,Zeppenfeld2012a,Jansen2013,Chervenkov2014,Barry2014,Lu2014}. Much of this attention has focused on diatomic molecules, despite unique possibilities for polyatomic molecules~\cite{Wall2013,Wall2015,Daussy1999,Tesch2002}. The latter possess additional rotational and vibrational degrees of freedom which could be used for various applications. For example, symmetric-top molecules have been suggested to be ideally suited to simulate quantum magnetism~\cite{Wall2013,Wall2015}. Moreover, precision tests of physics based on chirality require molecules with at least four atoms~\cite{Daussy1999}. In addition, single large molecules have been suggested for the realization of an entire quantum computer, using different vibrational modes to encode individual qubits~\cite{Tesch2002}. Last but not least, the high vapor pressure for many polyatomic molecule species, even at room temperature, allows the efficient generation of high-density initial ensembles~\cite{Junglen2004}.

A key challenge for obtaining cold and ultracold molecular ensembles has been gaining and maintaining control of the internal molecular state. While this is true for molecules in general, it is particularly problematic for larger, polyatomic molecules. Thus, even for the relatively light molecule CH$_3$F discussed here, several thousand rotational states are populated at room temperature. For larger molecules, a huge number of states is populated even at liquid-helium temperatures. Gaining quantum-state control of such molecules requires some form of internal-state cooling. While internal-state cooling has been demonstrated for bialkali dimers~\cite{Viteau2008,Manai2012,Wakim2012} as well as for a number of diatomic molecular ions~\cite{Staanum2010,Schneider2010,Rellergert2013,Hansen2014,Lien2014}, its implementation for polyatomic molecules is lacking.

In this Letter, we demonstrate comprehensive internal-state control of the polyatomic molecule methyl fluoride (CH$_3$F). In a two-step process, molecules in $16$ rotational $M$-sublevels in the lowest four rotational $J$ states in the $|K|{=}3$ manifold are optically pumped into a single rotational $M$-sublevel  ($J,K,M$ being the usual symmetric-top rotational quantum numbers). As a first step, we demonstrate rotational-state cooling (RSC) by optically pumping molecules in the states $J{=}5$ and $6$ into the state $J{=}4$, with minimal control of the $M$-sublevel. Molecules in the states $J{=}3$ and $4$ can then be motionally cooled via optoelectrical Sisyphus cooling~\cite{Zeppenfeld2009,Zeppenfeld2012a}, which can be integrated seamlessly with our RSC. This also eliminates all molecules in uncooled states from our trap. As a second step, an $M$-sublevel-dependent optical pumping transfers the molecules into the single state $J{=}4$, $K{=}3$, $M{=}4$.

\begin{figure}[t]
\centering
\includegraphics{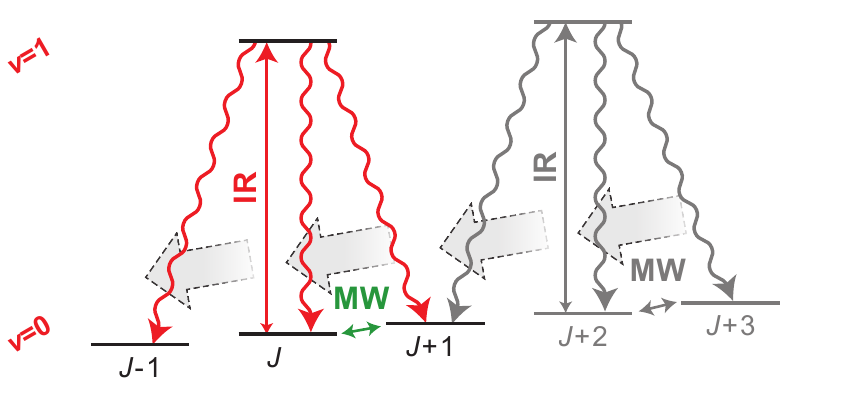}
\caption{Simplified scheme for rotational-state cooling (RSC). Optical pumping via a vibrational excited state with infrared (IR) and microwave (MW) radiation is used to accumulate population in a lower rotational state. 
}
\label{fig:Fig1}
\end{figure}

Our RSC scheme is based on optical pumping via excitation of a vibrational mode using a scheme related to those demonstrated for diatomic molecular ions~\cite{Staanum2010,Schneider2010}. The underlying idea of our scheme is shown in Fig.~\ref{fig:Fig1}. Driving a $\Delta J{=}0$ vibrational transition from a given rotational state $J$ results in spontaneous decay to the states $J{\pm}1$. By coupling the states $J$ and $J{+}1$ with microwaves, the entire population in the states $J$ and $J{\pm}1$ accumulates in the dark state $J{-}1$. This scheme can easily be extended to incorporate larger sets of initial states by, e.g., applying the same couplings to subsequent pairs of rotational states, as sketched in Fig.~\ref{fig:Fig1}.
Alternatively, a single vibrational transition might be used if all states $J'\ge J$ are coupled with microwave radiation. In principle, the dark state of such a scheme can be freely chosen by picking a suitable microwave coupling.

State manipulation takes place in our homogeneous-field electric trap (volume${\approx}\unit[2]{cm^3}$)~\cite{Englert2011,Zeppenfeld2013}. The narrow electric-field distribution of the trap~\cite{Gloeckner2015} is essential for RSC, as this allows us to spectrally resolve all relevant transitions, with minimal Stark broadening. In addition, the long trap lifetime of up to $\unit[30]{s}$~\cite{Zeppenfeld2012a} is crucial for the implementation of an optical pumping scheme based on spontaneous decays of vibrational excitations with typical decay times of more than $\unit[10]{ms}$. Despite its slow decay rate, the use of a vibrational mode is favorable compared to electronic transitions due to strict selection rules. Moreover, electronic excitation for many polyatomic molecules leads to rapid fragmentation.

\begin{figure}[!pt]
\centering
\includegraphics{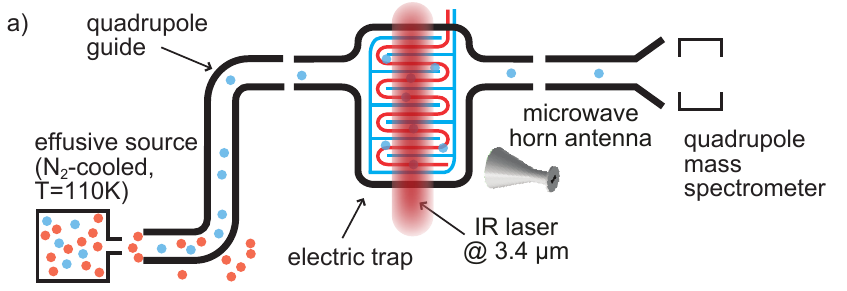}
\includegraphics{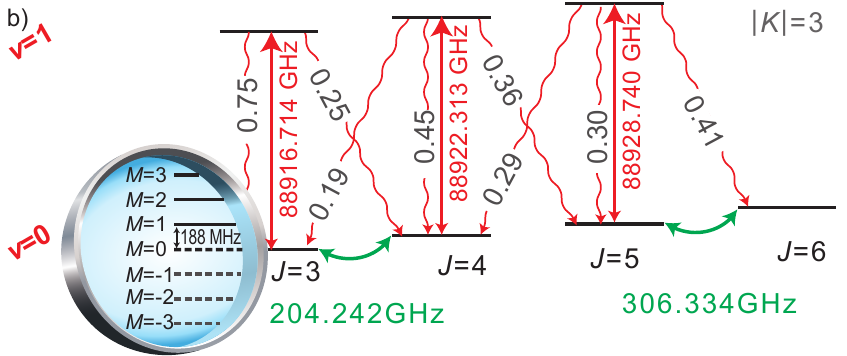}
\caption{Experimental setup and level scheme.
(a)~Experimental setup as described in the main text. 
(b)~Level scheme of CH$_3$F for the four lowest rotational states of the $|K|{=}3$ manifold. Transition frequencies are given for zero electric field. As indicated by the zoom, the rotational states split in an electric field according to the linear Stark effect. Only the low-field seeking states (positive $M$ values) are trapped with our DC electric trap. The splitting in the homogeneous-field region of our trap (\unit[815]{V/cm}) is \unit[188,113,75,54]{MHz} for $J{=}3,J{=}4,J{=}5,J{=}6$, respectively. Additional figures of the optical pumping schemes described in the main text are provided in~\cite{NoteSup}.}
\label{fig:Fig2}
\end{figure}

The trap is integrated in the experimental setup as shown in Fig.~\ref{fig:Fig2}(a). 
An experiment starts with loading molecules from a velocity-filtered thermal source ($T{\approx}\unit[110]{K}$) via an electric quadrupole guide~\cite{Junglen2004}. Subsequently, the ensemble is stored in the trap for manipulation. Light from an optical parametric oscillator at \unit[3.4]{\textmu m} is used to drive the parallel $v_1$ {C-H} stretch vibrational transition. Rotational levels are coupled by microwave radiation at \unit[200]{GHz} and \unit[300]{GHz} generated by amplifier-multiplier chains~\cite{Zeppenfeld2012a}.
Finally, molecules are guided to a quadrupole mass spectrometer for detection.
The experimental sequences differ in the time for trap loading $t_{l}$, storage $t_s$ and unloading $t_u$, which we detail as $t_{seq}{=}\left(t_l,t_s,t_u\right)$ together with the data. 

The rotational state distribution of the molecules loaded into our trap spreads over tens of rotational states~\cite{Gloeckner2015}. We label these states by the vibrational quantum number $v$ and symmetric-top rotational quantum numbers $J$, $K$, $M$ as $|v{;}J{,}{\mp} K{,}{\pm} M\rangle$ with $\mp K$ chosen positive. The lowest rotational states of the $|K|{=}3$ manifold have the highest population of about 17\% in $J{=}3,4$ and 9\% in $J{=}5{,}6$~\cite{Gloeckner2015}, with about a million molecules in each of these states~\cite{Zeppenfeld2012a}. We therefore have chosen these states to demonstrate RSC.

The population within a given set of rotational states is detected by state-selective removal of these molecules from the trap with MW and IR radiation at the end of the storage period as described in the supplemental material~\cite{NoteSup}  and in reference~\cite{Gloeckner2015}. Our present radiation sources allow for the detection of particular sets of rotational states, for example $J{=}3,4$ and $J{=}3{,}4{,}5{,}6$ with $|K|{=}3$ or the single rotational $M$-sublevels $|0{;}4,3,4\rangle$ and $|0{;}3,3,3\rangle$. 

The level scheme for the rotational states used in this Letter is shown in Fig.~\ref{fig:Fig2}(b). For the purpose of RSC and detection, it is required to drive the $\Delta J{=}0$ vibrational transitions shown simultaneously. We realized a quasi-simultaneous driving with a single laser source by suitably changing the frequency every \unit[20]{ms}. This is possible due to the use of $\Delta J{=}0$ excitations with\mbox{ - in general - }close lying transition frequencies.

\begin{figure}[!pt]
\centering
\includegraphics{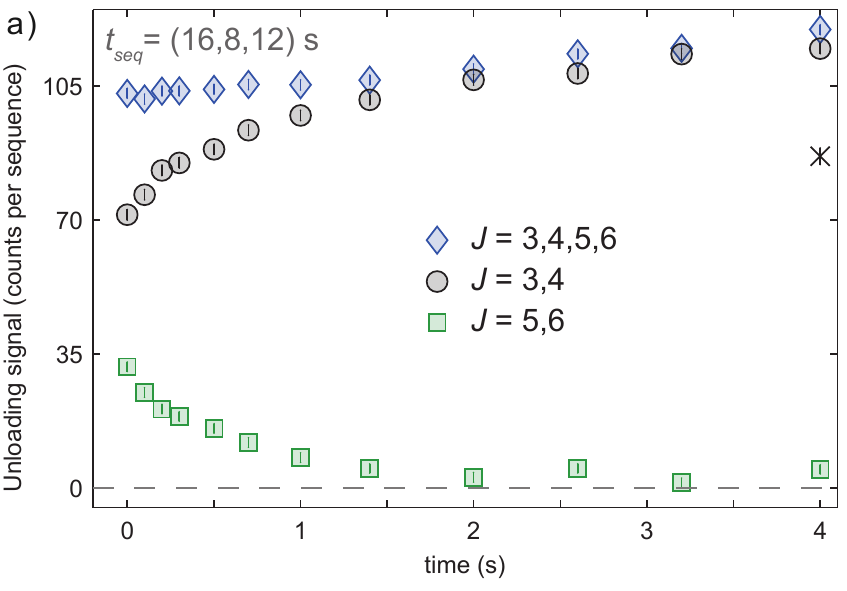}
\includegraphics{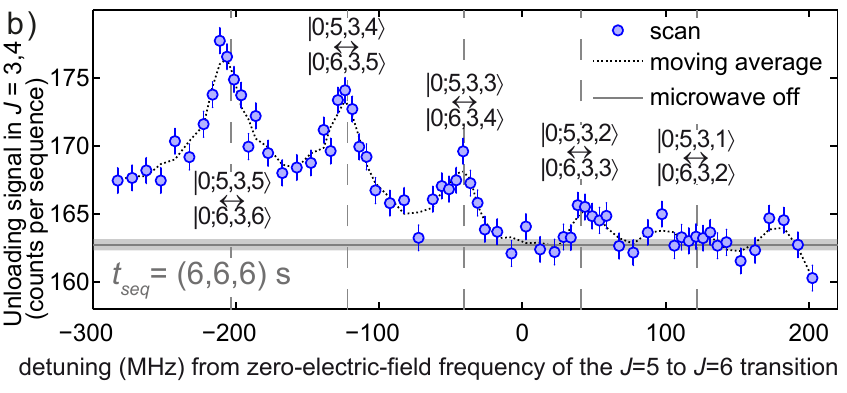}
\caption{ Results for RSC. We plot the rotational-state-discriminated unloading signal detected with the quadrupole mass spectrometer.   
(a)~A time-dependent measurement of RSC. The single black cross was measured while leaving the microwaves off (see main text).  
(b)~Contribution of single $M$-sublevels to RSC. Only a single MW frequency is applied and scanned across all relevant transitions coupling $J{=}5{,}6$ (dashed vertical lines show the calculated transition frequencies). 
To increase the resolution of the individual transitions, a higher electric field strength was used. This results in a larger separation of the $M$-substates and thus a larger splitting of the transition frequencies.
Vertical bars denote the $1\sigma$ statistical error.}
\label{fig:Fig3}
\end{figure}

To implement the RSC scheme shown in Fig.~\ref{fig:Fig1} we have chosen to transfer population from the rotational states $J{=}5{,}6$ to $J{=}4$. A $\Delta M{=}0$ vibrational transition from the $J{=}5$ state is driven together with the $|0{;}5,3,M\rangle \leftrightarrow |0{;}6,3,M{+}1\rangle$ ($M{>}0$) rotational transitions. Since the optical pumping scheme employs a $\Delta J{=}0$ vibrational transition, the Stark splittings of ground and excited state are almost identical. Consequently, we can address all $M$-sublevels simultaneously with one laser frequency. The selection rules for electric dipole transitions $\Delta J,\Delta M {=} 0,{\pm}1$ and $\Delta K {=} 0$ and the choice of infrared and microwave couplings  ensure that most molecules stay trapped while they are pumped from $|0{;}5,3,M\rangle$ and $|0{;}6,3,M\rangle$ to $|0{;}4{,}3{,}M^{\prime}\rangle$. These favorable selection rules and branching ratios are generic properties of symmetric top molecules.  

For proof of RSC, we present two measurements. First, RSC is applied for a varying amount of time. Fig.~\ref{fig:Fig3}(a) shows the resulting clear increase of population in $J{=}3,4$ together with a decrease of population in $J{=}5{,}6$ to almost zero. 
The signal for all four examined states $J{=}3{,}4{,}5{,}6$ shows a slight increase of molecule number for long pumping times, although a simple rate model suggests a loss of about 10\% of the molecules to untrapped states during RSC. The observed effect can be explained by different trap lifetimes for different states arising from an increase of the mean Stark shift in the course of optical pumping~\cite{NoteSup}. Fig.~\ref{fig:Fig3}(a) also gives the timescale of RSC which matches nicely with the rate model in the supplementary information. The relatively long timescale is due to the slow spontaneous decay rate from the excited vibrational state of approximately $15$\,Hz, combined with a probability of at most $1/3$ for a molecule to be in the excited vibrational state and a branching ratio of $0.29$ for a decay from the vibrational excited state to the states $|0{;}4{,}3{,}M\rangle$.

While Fig.~\ref{fig:Fig3}(a) shows an increase of molecules in the states $J{=}3{,}4$ closely tracking the decrease of molecules in $J{=}5{,}6$, it does not prove conclusively that the increased signal in $J{=}3{,}4$ originates from molecules initially in the states $J{=}5{,}6$. We therefore explore the effect of the microwaves coupling the states $J{=}5$ and $6$ on the RSC. Simply leaving these microwaves off leads to significantly less population in the states with $J{=}3{,}4$ for \unit[4]{s} of cooling, as shown by the cross in Fig.~\ref{fig:Fig3}(a). Since the microwaves only affect molecules in the states $J{=}5$ and $6$, the increase of population in $J{=}3{,}4$ over time clearly originates from molecules in the states $J{=}5{,}6$. Note that the increase of molecules in $J{=}3{,}4$ without microwaves compared to no cooling is consistent with the number of molecules expected to be transferred from $J{=}5$ to $J{=}4$ by the vibrational transition alone.

In a second measurement, we examine the effect of the microwave radiation coupling $J{=}5$ and $6$ on the population of molecules transferred to $J{=}3{,}4$ more closely and identify the contribution of individual $M$-sublevels in $J{=}6$. We therefore rotationally cool for \unit[2]{s} with only one microwave frequency applied and monitor the signal in $J{=}3{,}4$ while scanning this frequency. The measured spectrum is shown in Fig.~\ref{fig:Fig3}(b). The Stark-shifted transitions coupling the three highest $M$-sublevels can be clearly identified proving a transfer of molecules from $|0{;}6{,}3{,}M{=}4{,}5{,}6\rangle$ to $|0{;}4{,}3{,}M'{>}0\rangle$. Lower lying $M$-substates yield smaller or negligible peaks because these are less populated initially and the chance of losing the corresponding molecules to untrapped states during the pumping process is higher.

A key advantage of our RSC scheme is that it can be integrated straightforwardly with motional Sisyphus cooling and thus enables the simultaneous cooling of the internal and external degrees of freedom. Sisyphus cooling is expected to provide temperatures below \unit[1]{mK} for many molecule species and was first demonstrated with the same experimental apparatus in 2012~\cite{Zeppenfeld2012a}. In that work CH$_3$F molecules populating the states $J=3,4;|K|=3$ were motionally cooled from $T{\approx} \unit[0.4]{K}$ to ${\approx}\unit[30]{mK}$. 
As the main difference to RSC, kinetic energy is extracted by driving radio frequency transitions (\unit[3]{GHz} to \unit[0.4]{GHz}) from a highest $M$-substate to a lower $M$ in edge regions of the trap with high electric field. Molecules are optically pumped back to the initial, strongly trapped states by driving the $\Delta J{=}0,\Delta M{=}{+}1$ vibrational transitions from $J{=}3$ in the homogeneous-electric-field region (with lower electric fields). Due to the spontaneous decay to $J{=}4$ we additionally couple the $|0{;}3{,}3{,}M\rangle {\leftrightarrow} |0{;}4{,}3{,}M{+}1\rangle$ ($M{>}0$) rotational states with MW radiation to obtain a closed level scheme. In this way, a large fraction of a molecule's kinetic energy can be extracted in each cooling cycle and several repetitions are possible with minimal losses to untrapped states.

We demonstrate the concurrent motional and rotational cooling by adding the RSC to the Sisyphus cooling sequence. The addition of RSC results in an equally cold, but larger ensemble of molecules. By applying RSC only during trap loading and the subsequent second, almost all molecules entering the trap in the states $J{=}5{,}6$ are pumped to $J{=}4$. As expected from the initial rotational-state distribution, this increases the signal of cold molecules by 49(4)\,\%. By continuing the optical pumping during the entire motional-cooling sequence we increase the signal by 76(4)\,\%. This closes two loss channels of the Sisyphus scheme specific to $\rm CH_3F$ which transfer population to the $J{=}5{;}|K|{=}3$ state. These loss channels are the presence of a Fermi resonance in the vibrationally excited state~\cite{Zeppenfeld2012a} and an excitation of a distinct vibrational mode by blackbody radiation (estimated excitation rate \unit[0.08]{Hz})~\cite{Zeppenfeld2013}.

The RSC demonstrated thus far leaves molecules in a number of $M$-sublevels in the states $J{=}3{,}4$, or, combined with Sisyphus cooling, in the two states $|0{;}3{,}3{,}3\rangle$ and $|0{;}4{,}3{,}4\rangle$~\cite{Zeppenfeld2012a}. As a final experiment, we now show the preparation of the cooled molecular sample in the single rotational $M$-substate $|0{;}4{,}3{,}4\rangle$. This can be achieved by appending another optical pumping scheme to the experimental sequence for combined rotational and motional cooling.  Specifically, we drive the vibrational transitions $\Delta M{=}0$ from $J{=}3$ and $\Delta M{=}{+}1$ from $J{=}4$ together with the RSC described above ($J{=}5{,}6{\rightarrow}J{=}4$). This leads to $|0{;}4{,}3{,}4\rangle$ being the only dark state in the manifold $J{=}3{,}4{,}5{,}6, |K|{=}3$. To characterize this single-state preparation (SSP) process, we first evenly distribute the population of the cooled molecular ensemble among the states $|0{;}3{,}3{,}3\rangle$ and $|0{;}4{,}3{,}4\rangle$ by coupling the two with MW. The molecules in $|0{;}4{,}3{,}4\rangle$ are then measured with or without \unit[2]{s} of SSP applied. We find that SSP increases the absolute number of molecules populating the state $|0{;}4{,}3{,}4\rangle$ by 70 (3)\,\%, with this number also yielding the optical-pumping efficiency. A slight modification of the scheme also allows pumping from $|0{;}4{,}3{,}4\rangle$ to $|0{;}3{,}3{,}3\rangle$, specifically by driving a $\Delta M{=}{+}1$ transition from $J{=}3$ and a $\Delta M{=}0$ transition from $J{=}4$. This yields a transfer efficiency of 50(5)\,\%, where the reduced value is a consequence of less favorable branching ratios. The presented experiments show that we can significantly increase the population in a highest $M$-sublevel of choice. We estimate~\cite{Zeppenfeld2012a} that finally about $10^6$ cooled molecules populate the target state. 

\begin{figure}[tbp]
\centering
\includegraphics{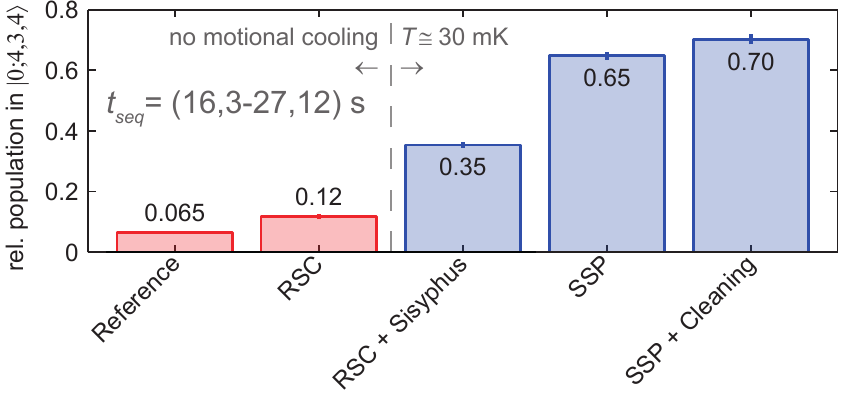}
\caption{Relative population of all trapped molecules in the $|0{;}4;3;4\rangle$ state for the following experimental sequences:
Reference: no state manipulation $(t_s{=}3\,\rm{s})$;
RSC: rotational cooling as presented in Fig.~\ref{fig:Fig3} $(t_s{=}6\,\rm{s})$;
RSC + Sisyphus: motional cooling combined with RSC $(t_s{=}23\,\rm{s})$;
SSP: the previous sequence with subsequent optical pumping to $|0{;}4{,}3{,}4\rangle$ $(t_s{=}25\,\rm{s})$;
SSP + cleaning: in addition to SSP removal of all states except for the $|0{;}4{,}3{,}4\rangle$ state from the trap $(t_s{=}27\,\rm{s})$. 
The internal-state purity increases to 70\,\%, while the mean kinetic energy is reduced by more than an order of magnitude.
The vertical solid lines represent the $1\sigma$ statistical error (which is smaller than the edge of the bar if not visible).
The values for uncooled ensembles overestimate the population relative to the other sequences due to an imperfect state detection~\cite{Gloeckner2015}.
}
\label{fig:Fig4}
\end{figure}

All previous measurements examined optical pumping in the context of increasing the absolute number of molecules in a particular state or set of states. In the final part of this Letter we study the state purity of the molecular ensemble. The measured relative populations of $|0{;}4{,}3{,}4\rangle$ (normalized to the total trap signal) resulting from different preparation procedures are presented and explained in Fig.~\ref{fig:Fig4}. Mainly two processes enhance the rotational-state purity of the trapped sample of molecules in the course of the shown experiments: apart from optical pumping, which is specifically applied for this reason, filtering occurs as a side effect of motional Sisyphus cooling. During motional cooling, molecules initially populating states outside the closed cycling scheme consisting of $|0{;}3{,}3{,}M\rangle$, $|0{;}4{,}3{,}M\rangle$ and $|1{;}3{,}3{,}M\rangle$ are coupled to untrapped states with a slow rate~\cite{Zeppenfeld2012a}. Consequently, translational cooling alone leads to a significant increase of purity, since afterwards almost all molecules populate the states $|0{;}3{,}3{,}3\rangle$ and $|0{;}4{,}3{,}4\rangle$. However, deliberate optical pumping to the target state $|0{;}4{,}3{,}4\rangle$ (SSP) yields the greatest effect on state purity as can be expected from the previous discussions. The relative occupation is boosted a bit further by removing all molecules still populating states other than the dark state of the SSP scheme (so-called cleaning).

In total, we augment the population of a single $M$-sublevel from initially ${\sim} 6$\,\% to at least 70\,\% while the molecular ensemble is motionally cooled by more than an order of magnitude. 
In fact, we even expect ${\sim} 90$\,\% of the molecules to populate this particular rotational state, as our detection method underestimates the real occupation by roughly 20\,\%~\cite{Gloeckner2015}.
The achieved purity is mainly limited by the above discussed blackbody radiation, which removes molecules from the target state during application of SSP and cleaning with a rate of \unit[0.08]{Hz}. This limitation is of technical nature and can be eliminated by cooling down the electric trap or choosing a molecule species not suffering from significant blackbody-radiation losses.

By applying the extremely useful technique of optical pumping to trapped, neutral, polyatomic molecules we demonstrate unprecedented control of this system. Starting from many initial states and pumping into a single final state we produce a pure ensemble of cooled molecules making use of only a single laser and microwave radiation. The final rotational state can in principle be chosen freely. Our scheme is based on generic properties common to many molecule species and should therefore be applicable to a whole class of trappable molecules. Thus, as a next step, we plan on internally cooling heavier molecules which populate numerous internal states even at cryogenic temperatures. This extends the number of species accessible to motional Sisyphus cooling. Comprehensive internal-state control also gives new prospects for high precision measurements or state resolved collision studies. Finally, extending our scheme to pump into the absolute ground state~\cite{Watson1971} creates favorable conditions for the investigation of sympathetic or evaporative cooling of polyatomic molecules.

\clearpage
\newpage

\renewcommand{\thefigure}{S\arabic{figure}}
\setcounter{figure}{0}

\section*{Supplemental Material}

\subsection{Rotational-state detection}

In this section of the supplementary material, we briefly summarize the main aspects of our rotational state detection which is in detail described in reference~\cite{Gloeckner2015}. We detect the population of a particular set of states by transferring the population in these states to untrapped states. These molecules are hence removed from the trapped ensemble.

\begin{figure}[!pb]
\centering
\includegraphics{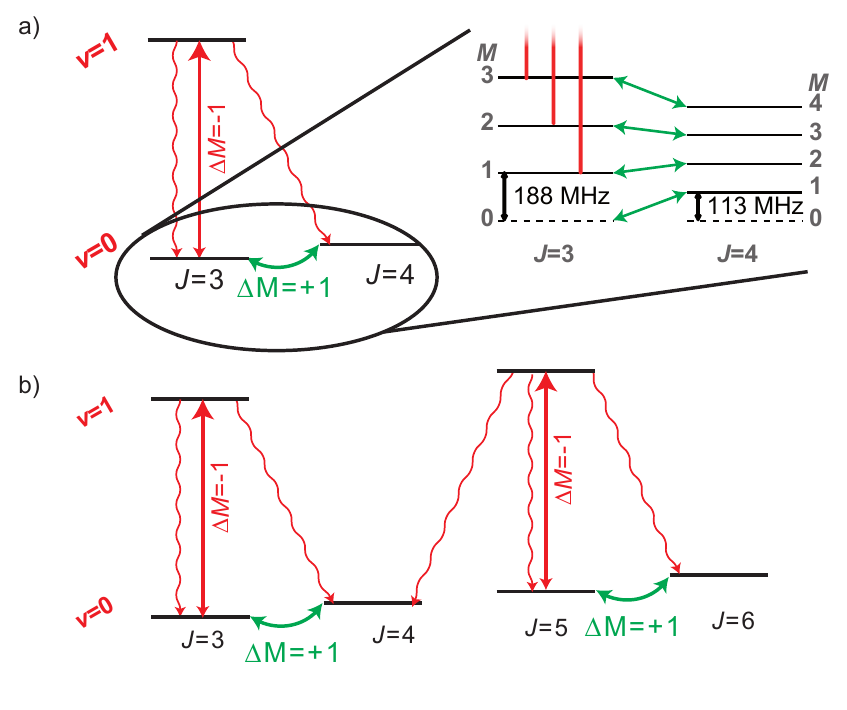}
\caption{Scheme for rotational state sensitive depletion.
(a)~IR depletion (IRD) for molecules populating the rotational states $J=3,4$. The inset shows the coupling of the individual $M$-sublevels of the rotational states $J=3$ and $J=4$ with MW. 
(b)~IRD for molecules populating the rotational states $J=3,4,5,6$.}
\label{fig:depl}
\end{figure}

Figure~\ref{fig:depl}(a) shows the IR and MW radiation used to state selectively remove (or deplete) molecules in the states $J=3,4$. We drive $\Delta M=-1$ vibrational transitions from the state $J=3$ and use the subsequent spontaneous decay to transfer the population to lower lying $M$-substates until they reach the untrapped state $M=0$. Due to spontaneous decay to the neighboring $J=4$ states we in addition drive the $|0{;}3,3,M\rangle \leftrightarrow |0{;}4,3,M+1\rangle$ rotational transitions. We thus deplete all molecules populating the set of states $J=3,4$.

To deplete the population in all four rotational states $J=3,4,5,6$ we in addition drive the $\Delta M=-1$ vibrational transitions from the state $J=5$ and couple the $|0{;}5,3,M\rangle \leftrightarrow |0{;}6,3,M+1\rangle$ rotational transitions with MW as shown in Fig.~\ref{fig:depl}(b). Due to the spontaneous decay to the rotational states $J\pm1$ it is not possible to deplete only the rotational states $J=5,6$ with this scheme. However, a state selected signal of molecules populating the set of states $J=5,6$ is given by the difference of signals measured while depleting $J=3,4,5,6$ or $J=3,4$. The sets of states which can be detected clearly depend strongly on which microwave sources are available to couple neighboring $J$ states.

To detect the population of the single $M$-sublevel $|0{;}4{,}3{,}4\rangle$ we use the scheme shown in Fig.~\ref{fig:depl}(a) without coupling the $|0{;}3{,}3{,}3\rangle \leftrightarrow |0{;}4,3,4\rangle$ states. Thus, the single $M$-sublevel $|0{;}4{,}3{,}4\rangle$ is the only one within $J=3,4$ which is not depleted. A comparison to the depletion signal of all substates in $J=3,4$ gives the population in the single $M$-sublevel $|0{;}4{,}3{,}4\rangle$. The detection of the population in the single $M$-state $|0{;}3{,}3{,}3\rangle$ works similarly. In contrast to the detection of $|0{;}4{,}3{,}4\rangle$, the vibrational transition is here driven from the $J=4$ rotational state. As this leads to spontaneous decay to states with $J=5$, the population in states with $J=5,6$ is additionally removed by coupling all $M$-substates with MW to untrapped states.

\subsection{Rate Model of RSC}

\begin{figure}[!pt]
\centering
\includegraphics{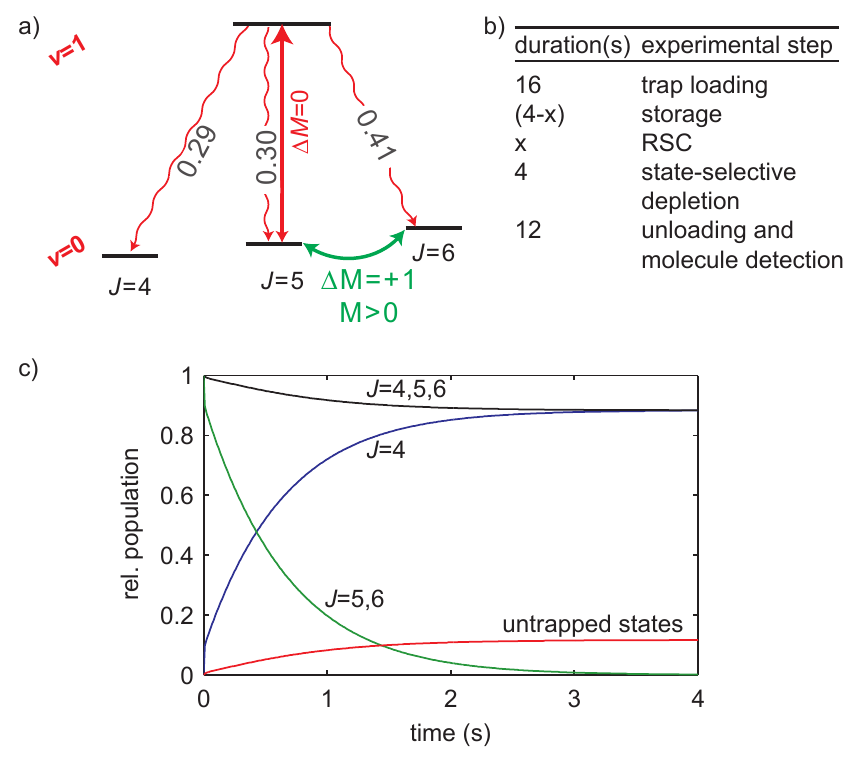}
\caption{Rotational-state cooling (RSC).
(a)~Scheme for RSC. 
(b)~Experimental sequence.
(c)~Results of a rate model for comparison with the measurement shown in the main Letter. Note that the state $J=3$ is not included in the simulation as it is not part of the optical pumping scheme. However, our current detection scheme does not allow to measure the population of $J=4$ alone. Therefore, the measurement presented in the Letter tracks the population of $J=3,4$ together.}
\label{fig:RSC}
\end{figure}

In Fig.~3(a) of our Letter we present the results of a time-dependent measurement of RSC. In this section of the supplementary material we discuss the results of a simple rate model and its limitations towards a comparison with the experimental data. RSC is experimentally implemented according to the scheme in Fig.~\ref{fig:RSC}(a). The changes in population caused by RSC can be calculated with a rate model. The rate model we use here simulates the driving of the MW and IR transitions needed for RSC by rate coefficients between appropriate states and the spontaneous decay from the vibrational excited state with a decay rate of \unit[15]{Hz}. The initial state distribution consists of all trappable $M$-substates in $J=5,6, |K|=3$. The distribution over the $M$-sublevels represents the conditions of our velocity filtered source~\cite{Gloeckner2015}. More detailed information about the rate model are given in reference~\cite{Gloeckner2015}.

The result is shown in Fig.~\ref{fig:RSC}(c). The time scale of the process matches the experimental results. In both cases, the population of the states with $J=5,6$ decreases by a factor of two within about \unit[0.4]{s}.  In addition, the rate model suggests about ten percent loss to untrapped states. However, we measure an increase of molecules in the total set of states $J=3,4,5,6$. The reasons for this discrepancy are described in the following. 

To understand the outcome of our measurement we first have to consider of the whole experimental sequence shown Fig.~\ref{fig:RSC}(b). In particular, the total time for storage is identical for all data points to minimize the influence of trap losses. As population transfer caused by blackbody radiation is present during that time, RSC is applied at the end of the storage time. We thus minimize repopulation of the states $J=5,6$ before detection. 

The order of the experimental sequence is the main reason for the measured increase of molecules in the total set of states $J=3,4,5,6$:
RSC on average increases the Stark shift of the molecule ($\delta_{\textrm{Stark}} \propto \frac{MK}{J(J+1)}$). For example, a spontaneous decay from the excited $|1{;}5{,}3{,}5\rangle$ state to $J=4$ has to end up in the $|0{;}4{,}3{,}4\rangle$ state, where the Stark shift is a factor of 1.2 higher. The excited $|1{;}5{,}3{,}3\rangle$ state can decay to the $|0{;}4{,}3{,}M{=}2,3,4\rangle$ states. The lowest possible $M$-sublevel $|0{;}4{,}3{,}2\rangle$ has the same Stark shift as $|1{;}5{,}3{,}3\rangle$. Therefore the molecules, again, end up in rotational states with on average higher Stark shifts. As the lifetime in our trap increases with the Stark shift of the molecule~\cite{Englert2011}, we expect to measure an increase in signal. This effect is particularly visible by comparing the results measured at \unit[2]{s} and \unit[4]{s}. Already after about \unit[2]{s} almost all molecules from the states $J=5,6$ were pumped to $J=3,4$ (or lost to untrapped states), but the total signal still slightly increases for longer RSC times. The molecules have been pumped to higher Stark shifts earlier in the experimental sequence and hence the trap losses are slightly reduced. 

Another experimental subtlety is different detection efficiencies in different internal states which also might hide losses. As discussed in reference~\cite{Gloeckner2015}, the detection efficiency of non manipulated molecules depends on the ratio of temperature and Stark shift. However, RSC slightly changes the stark shift possibly resulting in a slightly different detection efficiency. Both effects, the change in the molecule's lifetime and the possible different detection efficiency are hard to be included in the rate model but the outcome of our experiment can qualitatively be understood by the above considerations.

The loss of about 10\,\% of the molecules could be a limitation for extending the scheme to incorporate higher rotational states. However, instead of driving a $\Delta M=0$ IR transition one could first use a $\Delta M=+1$ transition which accumulates the population in the highest $M$-substates and then use the $\Delta M=0$ IR transition to address these states. For large sets of states even switching several times between a $\Delta M=0$ and $\Delta M=+1$ vibrational transition should be useful. In this way, the loss to untrapped states is strongly suppressed.

\subsection{Level schemes for optical pumping}

\begin{figure}[!pb]
\centering
\includegraphics{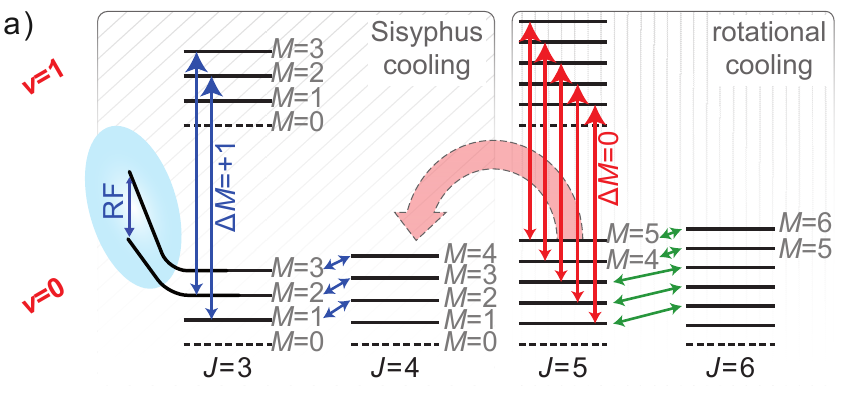}
\caption{Scheme for the combination of rotational and Sisyphus cooling. The right panel shows the IR and MW radiation needed for RSC as in Fig.~\ref{fig:RSC}(a). MW, IR and RF radiation employed for Sisyphus cooling are shown in the left panel. The RF transitions between the $M$-sublevels are driven in the high-field region of the trap as indicated in the blue ellipse for the two highest sublevels of $J=3$~\cite{Zeppenfeld2012a}.}
\label{fig:Comb}
\end{figure}

\begin{figure}[!pt]
\centering
\includegraphics{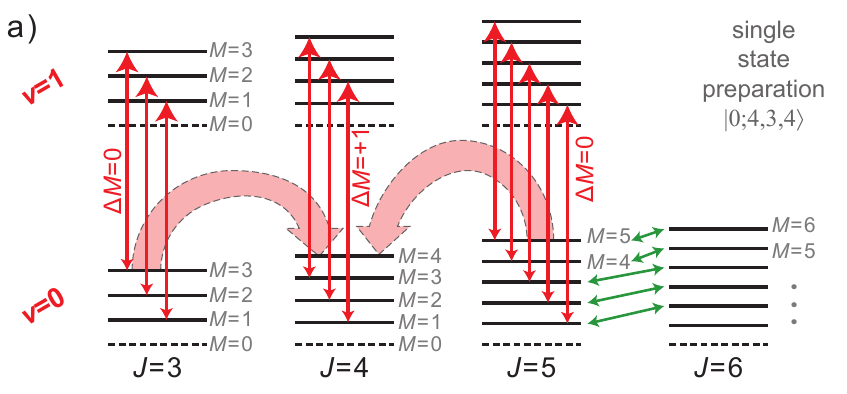}
\includegraphics{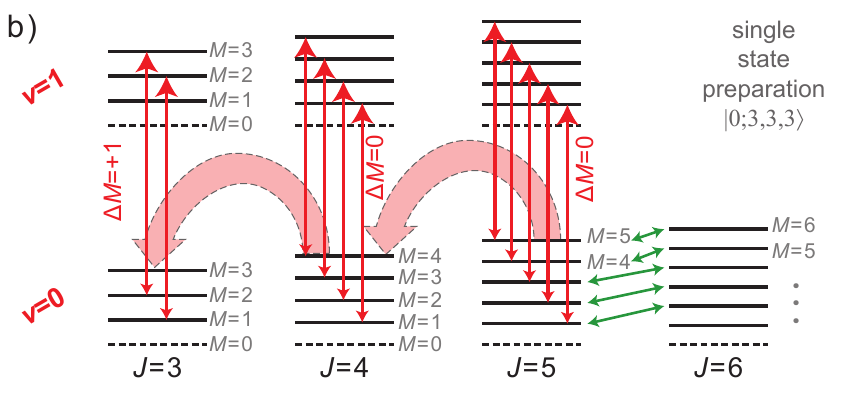}
\caption{Schemes for the single $M$-substate preparation in the states $|0{;}4{,}3{,}4\rangle$ (a) and $|0{;}3{,}3{,}3\rangle$(b)}
\label{fig:SSP}
\end{figure}

To simplify the understanding of the optical pumping schemes which are detailed in the Letter, we here append figures of the corresponding level schemes. First, Fig.~\ref{fig:Comb} shows the needed radiation for RSC (right hand side only) as well as the combination of Sisyphus cooling and RSC (whole figure). Second, Fig.~\ref{fig:SSP} gives the schemes for the preparation of the population in the single $M$-substates $|0{;}4{,}3{,}4\rangle$ and $|0{;}3{,}3{,}3\rangle$.

\begin{thebibliography}{28}%
\makeatletter
\providecommand \@ifxundefined [1]{%
 \@ifx{#1\undefined}
}%
\providecommand \@ifnum [1]{%
 \ifnum #1\expandafter \@firstoftwo
 \else \expandafter \@secondoftwo
 \fi
}%
\providecommand \@ifx [1]{%
 \ifx #1\expandafter \@firstoftwo
 \else \expandafter \@secondoftwo
 \fi
}%
\providecommand \natexlab [1]{#1}%
\providecommand \enquote  [1]{``#1''}%
\providecommand \bibnamefont  [1]{#1}%
\providecommand \bibfnamefont [1]{#1}%
\providecommand \citenamefont [1]{#1}%
\providecommand \href@noop [0]{\@secondoftwo}%
\providecommand \href [0]{\begingroup \@sanitize@url \@href}%
\providecommand \@href[1]{\@@startlink{#1}\@@href}%
\providecommand \@@href[1]{\endgroup#1\@@endlink}%
\providecommand \@sanitize@url [0]{\catcode `\\12\catcode `\$12\catcode
  `\&12\catcode `\#12\catcode `\^12\catcode `\_12\catcode `\%12\relax}%
\providecommand \@@startlink[1]{}%
\providecommand \@@endlink[0]{}%
\providecommand \url  [0]{\begingroup\@sanitize@url \@url }%
\providecommand \@url [1]{\endgroup\@href {#1}{\urlprefix }}%
\providecommand \urlprefix  [0]{URL }%
\providecommand \Eprint [0]{\href }%
\providecommand \doibase [0]{http://dx.doi.org/}%
\providecommand \selectlanguage [0]{\@gobble}%
\providecommand \bibinfo  [0]{\@secondoftwo}%
\providecommand \bibfield  [0]{\@secondoftwo}%
\providecommand \translation [1]{[#1]}%
\providecommand \BibitemOpen [0]{}%
\providecommand \bibitemStop [0]{}%
\providecommand \bibitemNoStop [0]{.\EOS\space}%
\providecommand \EOS [0]{\spacefactor3000\relax}%
\providecommand \BibitemShut  [1]{\csname bibitem#1\endcsname}%
\let\auto@bib@innerbib\@empty
\bibitem [{\citenamefont {Andr\'{e}}\ \emph {et~al.}(2006)\citenamefont
  {Andr\'{e}}, \citenamefont {DeMille}, \citenamefont {Doyle}, \citenamefont
  {Lukin}, \citenamefont {Maxwell}, \citenamefont {Rabl}, \citenamefont
  {Schoelkopf},\ and\ \citenamefont {Zoller}}]{ANDRE2006}%
  \BibitemOpen
  \bibfield  {author} {\bibinfo {author} {\bibfnamefont {A.}~\bibnamefont
  {Andr\'{e}}}, \bibinfo {author} {\bibfnamefont {D.}~\bibnamefont {DeMille}},
  \bibinfo {author} {\bibfnamefont {J.~M.}\ \bibnamefont {Doyle}}, \bibinfo
  {author} {\bibfnamefont {M.~D.}\ \bibnamefont {Lukin}}, \bibinfo {author}
  {\bibfnamefont {S.~E.}\ \bibnamefont {Maxwell}}, \bibinfo {author}
  {\bibfnamefont {P.}~\bibnamefont {Rabl}}, \bibinfo {author} {\bibfnamefont
  {R.~J.}\ \bibnamefont {Schoelkopf}}, \ and\ \bibinfo {author} {\bibfnamefont
  {P.}~\bibnamefont {Zoller}},\ }\bibfield  {title} {\enquote {\bibinfo {title}
  {{A coherent all-electrical interface between polar molecules and mesoscopic
  superconducting resonators}},}\ }\href {\doibase 10.1038/nphys386} {\bibfield
   {journal} {\bibinfo  {journal} {Nat. Phys.}\ }\textbf {\bibinfo {volume}
  {2}},\ \bibinfo {pages} {636} (\bibinfo {year} {2006})}\BibitemShut {NoStop}%
\bibitem [{\citenamefont {Micheli}\ \emph {et~al.}(2006)\citenamefont
  {Micheli}, \citenamefont {Brennen},\ and\ \citenamefont
  {Zoller}}]{Micheli2006}%
  \BibitemOpen
  \bibfield  {author} {\bibinfo {author} {\bibfnamefont {A.}~\bibnamefont
  {Micheli}}, \bibinfo {author} {\bibfnamefont {G.~K.}\ \bibnamefont
  {Brennen}}, \ and\ \bibinfo {author} {\bibfnamefont {P.}~\bibnamefont
  {Zoller}},\ }\bibfield  {title} {\enquote {\bibinfo {title} {{A toolbox for
  lattice-spin models with polar molecules}},}\ }\href {\doibase
  10.1038/nphys287} {\bibfield  {journal} {\bibinfo  {journal} {Nat. Phys.}\
  }\textbf {\bibinfo {volume} {2}},\ \bibinfo {pages} {341} (\bibinfo {year}
  {2006})}\BibitemShut {NoStop}%
\bibitem [{\citenamefont {Hudson}\ \emph {et~al.}(2011)\citenamefont {Hudson},
  \citenamefont {Kara}, \citenamefont {Smallman}, \citenamefont {Sauer},
  \citenamefont {Tarbutt},\ and\ \citenamefont {Hinds}}]{Hudson2011}%
  \BibitemOpen
  \bibfield  {author} {\bibinfo {author} {\bibfnamefont {J.~J.}\ \bibnamefont
  {Hudson}}, \bibinfo {author} {\bibfnamefont {D.~M.}\ \bibnamefont {Kara}},
  \bibinfo {author} {\bibfnamefont {I.~J.}\ \bibnamefont {Smallman}}, \bibinfo
  {author} {\bibfnamefont {B.~E.}\ \bibnamefont {Sauer}}, \bibinfo {author}
  {\bibfnamefont {M.~R.}\ \bibnamefont {Tarbutt}}, \ and\ \bibinfo {author}
  {\bibfnamefont {E.~A.}\ \bibnamefont {Hinds}},\ }\bibfield  {title} {\enquote
  {\bibinfo {title} {{Improved measurement of the shape of the electron.}}}\
  }\href {\doibase 10.1038/nature10104} {\bibfield  {journal} {\bibinfo
  {journal} {Nature}\ }\textbf {\bibinfo {volume} {473}},\ \bibinfo {pages}
  {493} (\bibinfo {year} {2011})}\BibitemShut {NoStop}%
\bibitem [{\citenamefont {Lemeshko}\ \emph {et~al.}(2013)\citenamefont
  {Lemeshko}, \citenamefont {Krems}, \citenamefont {Doyle},\ and\ \citenamefont
  {Kais}}]{Lemeshko2013a}%
  \BibitemOpen
  \bibfield  {author} {\bibinfo {author} {\bibfnamefont {M.}~\bibnamefont
  {Lemeshko}}, \bibinfo {author} {\bibfnamefont {R.~V.}\ \bibnamefont {Krems}},
  \bibinfo {author} {\bibfnamefont {J.~M.}\ \bibnamefont {Doyle}}, \ and\
  \bibinfo {author} {\bibfnamefont {S.}~\bibnamefont {Kais}},\ }\bibfield
  {title} {\enquote {\bibinfo {title} {{Manipulation of molecules with
  electromagnetic fields}},}\ }\href {\doibase 10.1080/00268976.2013.813595}
  {\bibfield  {journal} {\bibinfo  {journal} {Mol. Phys.}\ }\textbf {\bibinfo
  {volume} {111}},\ \bibinfo {pages} {1648} (\bibinfo {year}
  {2013})}\BibitemShut {NoStop}%
\bibitem [{\citenamefont {Ni}\ \emph {et~al.}(2008)\citenamefont {Ni},
  \citenamefont {Ospelkaus}, \citenamefont {de~Miranda}, \citenamefont {Pe'er},
  \citenamefont {Neyenhuis}, \citenamefont {Zirbel}, \citenamefont
  {Kotochigova}, \citenamefont {Julienne}, \citenamefont {Jin},\ and\
  \citenamefont {Ye}}]{Ni2008}%
  \BibitemOpen
  \bibfield  {author} {\bibinfo {author} {\bibfnamefont {K.-K.}\ \bibnamefont
  {Ni}}, \bibinfo {author} {\bibfnamefont {S.}~\bibnamefont {Ospelkaus}},
  \bibinfo {author} {\bibfnamefont {M.~H.~G.}\ \bibnamefont {de~Miranda}},
  \bibinfo {author} {\bibfnamefont {A.}~\bibnamefont {Pe'er}}, \bibinfo
  {author} {\bibfnamefont {B.}~\bibnamefont {Neyenhuis}}, \bibinfo {author}
  {\bibfnamefont {J.~J.}\ \bibnamefont {Zirbel}}, \bibinfo {author}
  {\bibfnamefont {S.}~\bibnamefont {Kotochigova}}, \bibinfo {author}
  {\bibfnamefont {P.~S.}\ \bibnamefont {Julienne}}, \bibinfo {author}
  {\bibfnamefont {D.~S.}\ \bibnamefont {Jin}}, \ and\ \bibinfo {author}
  {\bibfnamefont {J.}~\bibnamefont {Ye}},\ }\bibfield  {title} {\enquote
  {\bibinfo {title} {{A high phase-space-density gas of polar molecules.}}}\
  }\href {\doibase 10.1126/science.1163861} {\bibfield  {journal} {\bibinfo
  {journal} {Science}\ }\textbf {\bibinfo {volume} {322}},\ \bibinfo {pages}
  {231} (\bibinfo {year} {2008})}\BibitemShut {NoStop}%
\bibitem [{\citenamefont {Zeppenfeld}\ \emph {et~al.}(2012)\citenamefont
  {Zeppenfeld}, \citenamefont {Englert}, \citenamefont {Gl\"{o}ckner},
  \citenamefont {Prehn}, \citenamefont {Mielenz}, \citenamefont {Sommer},
  \citenamefont {van Buuren}, \citenamefont {Motsch},\ and\ \citenamefont
  {Rempe}}]{Zeppenfeld2012a}%
  \BibitemOpen
  \bibfield  {author} {\bibinfo {author} {\bibfnamefont {M.}~\bibnamefont
  {Zeppenfeld}}, \bibinfo {author} {\bibfnamefont {B.~G.~U.}\ \bibnamefont
  {Englert}}, \bibinfo {author} {\bibfnamefont {R.}~\bibnamefont
  {Gl\"{o}ckner}}, \bibinfo {author} {\bibfnamefont {A.}~\bibnamefont {Prehn}},
  \bibinfo {author} {\bibfnamefont {M.}~\bibnamefont {Mielenz}}, \bibinfo
  {author} {\bibfnamefont {C.}~\bibnamefont {Sommer}}, \bibinfo {author}
  {\bibfnamefont {L.~D.}\ \bibnamefont {van Buuren}}, \bibinfo {author}
  {\bibfnamefont {M.}~\bibnamefont {Motsch}}, \ and\ \bibinfo {author}
  {\bibfnamefont {G.}~\bibnamefont {Rempe}},\ }\bibfield  {title} {\enquote
  {\bibinfo {title} {{Sisyphus cooling of electrically trapped polyatomic
  molecules.}}}\ }\href {\doibase 10.1038/nature11595} {\bibfield  {journal}
  {\bibinfo  {journal} {Nature}\ }\textbf {\bibinfo {volume} {491}},\ \bibinfo
  {pages} {570} (\bibinfo {year} {2012})}\BibitemShut {NoStop}%
\bibitem [{\citenamefont {Jansen}\ \emph {et~al.}(2013)\citenamefont {Jansen},
  \citenamefont {Quintero-P\'{e}rez}, \citenamefont {Wall}, \citenamefont
  {van~den Berg}, \citenamefont {Hoekstra},\ and\ \citenamefont
  {Bethlem}}]{Jansen2013}%
  \BibitemOpen
  \bibfield  {author} {\bibinfo {author} {\bibfnamefont {P.}~\bibnamefont
  {Jansen}}, \bibinfo {author} {\bibfnamefont {M.}~\bibnamefont
  {Quintero-P\'{e}rez}}, \bibinfo {author} {\bibfnamefont {T.~E.}\ \bibnamefont
  {Wall}}, \bibinfo {author} {\bibfnamefont {J.~E.}\ \bibnamefont {van~den
  Berg}}, \bibinfo {author} {\bibfnamefont {S.}~\bibnamefont {Hoekstra}}, \
  and\ \bibinfo {author} {\bibfnamefont {H.~L.}\ \bibnamefont {Bethlem}},\
  }\bibfield  {title} {\enquote {\bibinfo {title} {{Deceleration and trapping
  of ammonia molecules in a traveling-wave decelerator}},}\ }\href {\doibase
  10.1103/PhysRevA.88.043424} {\bibfield  {journal} {\bibinfo  {journal} {Phys.
  Rev. A}\ }\textbf {\bibinfo {volume} {88}},\ \bibinfo {pages} {043424}
  (\bibinfo {year} {2013})}\BibitemShut {NoStop}%
\bibitem [{\citenamefont {Chervenkov}\ \emph {et~al.}(2014)\citenamefont
  {Chervenkov}, \citenamefont {Wu}, \citenamefont {Bayerl}, \citenamefont
  {Rohlfes}, \citenamefont {Gantner}, \citenamefont {Zeppenfeld},\ and\
  \citenamefont {Rempe}}]{Chervenkov2014}%
  \BibitemOpen
  \bibfield  {author} {\bibinfo {author} {\bibfnamefont {S.}~\bibnamefont
  {Chervenkov}}, \bibinfo {author} {\bibfnamefont {X.}~\bibnamefont {Wu}},
  \bibinfo {author} {\bibfnamefont {J.}~\bibnamefont {Bayerl}}, \bibinfo
  {author} {\bibfnamefont {A.}~\bibnamefont {Rohlfes}}, \bibinfo {author}
  {\bibfnamefont {T.}~\bibnamefont {Gantner}}, \bibinfo {author} {\bibfnamefont
  {M.}~\bibnamefont {Zeppenfeld}}, \ and\ \bibinfo {author} {\bibfnamefont
  {G.}~\bibnamefont {Rempe}},\ }\bibfield  {title} {\enquote {\bibinfo {title}
  {{Continuous Centrifuge Decelerator for Polar Molecules}},}\ }\href {\doibase
  10.1103/PhysRevLett.112.013001} {\bibfield  {journal} {\bibinfo  {journal}
  {Phys. Rev. Lett.}\ }\textbf {\bibinfo {volume} {112}},\ \bibinfo {pages}
  {013001} (\bibinfo {year} {2014})}\BibitemShut {NoStop}%
\bibitem [{\citenamefont {Barry}\ \emph {et~al.}(2014)\citenamefont {Barry},
  \citenamefont {McCarron}, \citenamefont {Norrgard}, \citenamefont
  {Steinecker},\ and\ \citenamefont {DeMille}}]{Barry2014}%
  \BibitemOpen
  \bibfield  {author} {\bibinfo {author} {\bibfnamefont {J.~F.}\ \bibnamefont
  {Barry}}, \bibinfo {author} {\bibfnamefont {D.~J.}\ \bibnamefont {McCarron}},
  \bibinfo {author} {\bibfnamefont {E.~B.}\ \bibnamefont {Norrgard}}, \bibinfo
  {author} {\bibfnamefont {M.~H.}\ \bibnamefont {Steinecker}}, \ and\ \bibinfo
  {author} {\bibfnamefont {D.}~\bibnamefont {DeMille}},\ }\bibfield  {title}
  {\enquote {\bibinfo {title} {{Magneto-optical trapping of a diatomic
  molecule}},}\ }\href {\doibase 10.1038/nature13634} {\bibfield  {journal}
  {\bibinfo  {journal} {Nature}\ }\textbf {\bibinfo {volume} {512}},\ \bibinfo
  {pages} {286} (\bibinfo {year} {2014})}\BibitemShut {NoStop}%
\bibitem [{\citenamefont {Lu}\ \emph {et~al.}(2014)\citenamefont {Lu},
  \citenamefont {Kozyryev}, \citenamefont {Hemmerling}, \citenamefont
  {Piskorski},\ and\ \citenamefont {Doyle}}]{Lu2014}%
  \BibitemOpen
  \bibfield  {author} {\bibinfo {author} {\bibfnamefont {H.-I.}\ \bibnamefont
  {Lu}}, \bibinfo {author} {\bibfnamefont {I.}~\bibnamefont {Kozyryev}},
  \bibinfo {author} {\bibfnamefont {B.}~\bibnamefont {Hemmerling}}, \bibinfo
  {author} {\bibfnamefont {J.}~\bibnamefont {Piskorski}}, \ and\ \bibinfo
  {author} {\bibfnamefont {J.~M.}\ \bibnamefont {Doyle}},\ }\bibfield  {title}
  {\enquote {\bibinfo {title} {{Magnetic Trapping of Molecules via Optical
  Loading and Magnetic Slowing}},}\ }\href {\doibase
  10.1103/PhysRevLett.112.113006} {\bibfield  {journal} {\bibinfo  {journal}
  {Phys. Rev. Lett.}\ }\textbf {\bibinfo {volume} {112}},\ \bibinfo {pages}
  {113006} (\bibinfo {year} {2014})}\BibitemShut {NoStop}%
\bibitem [{\citenamefont {Wall}\ \emph {et~al.}(2013)\citenamefont {Wall},
  \citenamefont {Maeda},\ and\ \citenamefont {Carr}}]{Wall2013}%
  \BibitemOpen
  \bibfield  {author} {\bibinfo {author} {\bibfnamefont {M.~L.}\ \bibnamefont
  {Wall}}, \bibinfo {author} {\bibfnamefont {K.}~\bibnamefont {Maeda}}, \ and\
  \bibinfo {author} {\bibfnamefont {L.~D.}\ \bibnamefont {Carr}},\ }\bibfield
  {title} {\enquote {\bibinfo {title} {{Simulating quantum magnets with
  symmetric top molecules}},}\ }\href {\doibase 10.1002/andp.201300105}
  {\bibfield  {journal} {\bibinfo  {journal} {Ann. Phys.}\ }\textbf {\bibinfo
  {volume} {525}},\ \bibinfo {pages} {845} (\bibinfo {year}
  {2013})}\BibitemShut {NoStop}%
\bibitem [{\citenamefont {Wall}\ \emph {et~al.}(2015)\citenamefont {Wall},
  \citenamefont {Maeda},\ and\ \citenamefont {Carr}}]{Wall2015}%
  \BibitemOpen
  \bibfield  {author} {\bibinfo {author} {\bibfnamefont {M.~L.}\ \bibnamefont
  {Wall}}, \bibinfo {author} {\bibfnamefont {K.}~\bibnamefont {Maeda}}, \ and\
  \bibinfo {author} {\bibfnamefont {L.~D.}\ \bibnamefont {Carr}},\ }\bibfield
  {title} {\enquote {\bibinfo {title} {{Realizing unconventional quantum
  magnetism with symmetric top molecules}},}\ }\href {\doibase
  10.1088/1367-2630/17/2/025001} {\bibfield  {journal} {\bibinfo  {journal}
  {New J. Phys.}\ }\textbf {\bibinfo {volume} {17}},\ \bibinfo {pages} {025001}
  (\bibinfo {year} {2015})}\BibitemShut {NoStop}%
\bibitem [{\citenamefont {Daussy}\ \emph {et~al.}(1999)\citenamefont {Daussy},
  \citenamefont {Marrel}, \citenamefont {Amy-Klein}, \citenamefont {Nguyen},
  \citenamefont {Bord\'{e}},\ and\ \citenamefont {Chardonnet}}]{Daussy1999}%
  \BibitemOpen
  \bibfield  {author} {\bibinfo {author} {\bibfnamefont {C.}~\bibnamefont
  {Daussy}}, \bibinfo {author} {\bibfnamefont {T.}~\bibnamefont {Marrel}},
  \bibinfo {author} {\bibfnamefont {A.}~\bibnamefont {Amy-Klein}}, \bibinfo
  {author} {\bibfnamefont {C.~T.}\ \bibnamefont {Nguyen}}, \bibinfo {author}
  {\bibfnamefont {C.~J.}\ \bibnamefont {Bord\'{e}}}, \ and\ \bibinfo {author}
  {\bibfnamefont {C.}~\bibnamefont {Chardonnet}},\ }\bibfield  {title}
  {\enquote {\bibinfo {title} {{Limit on the Parity Nonconserving Energy
  Difference between the Enantiomers of a Chiral Molecule by Laser
  Spectroscopy}},}\ }\href {\doibase 10.1103/PhysRevLett.83.1554} {\bibfield
  {journal} {\bibinfo  {journal} {Phys. Rev. Lett.}\ }\textbf {\bibinfo
  {volume} {83}},\ \bibinfo {pages} {1554} (\bibinfo {year}
  {1999})}\BibitemShut {NoStop}%
\bibitem [{\citenamefont {Tesch}\ and\ \citenamefont
  {de~Vivie-Riedle}(2002)}]{Tesch2002}%
  \BibitemOpen
  \bibfield  {author} {\bibinfo {author} {\bibfnamefont {C.~M.}\ \bibnamefont
  {Tesch}}\ and\ \bibinfo {author} {\bibfnamefont {R.}~\bibnamefont
  {de~Vivie-Riedle}},\ }\bibfield  {title} {\enquote {\bibinfo {title}
  {{Quantum Computation with Vibrationally Excited Molecules}},}\ }\href
  {\doibase 10.1103/PhysRevLett.89.157901} {\bibfield  {journal} {\bibinfo
  {journal} {Phys. Rev. Lett.}\ }\textbf {\bibinfo {volume} {89}},\ \bibinfo
  {pages} {157901} (\bibinfo {year} {2002})}\BibitemShut {NoStop}%
\bibitem [{\citenamefont {Junglen}\ \emph {et~al.}(2004)\citenamefont
  {Junglen}, \citenamefont {Rieger}, \citenamefont {Rangwala}, \citenamefont
  {Pinkse},\ and\ \citenamefont {Rempe}}]{Junglen2004}%
  \BibitemOpen
  \bibfield  {author} {\bibinfo {author} {\bibfnamefont {T.}~\bibnamefont
  {Junglen}}, \bibinfo {author} {\bibfnamefont {T.}~\bibnamefont {Rieger}},
  \bibinfo {author} {\bibfnamefont {S.~A.}\ \bibnamefont {Rangwala}}, \bibinfo
  {author} {\bibfnamefont {P.~W.~H.}\ \bibnamefont {Pinkse}}, \ and\ \bibinfo
  {author} {\bibfnamefont {G.}~\bibnamefont {Rempe}},\ }\bibfield  {title}
  {\enquote {\bibinfo {title} {{Slow ammonia molecules in an electrostatic
  quadrupole guide}},}\ }\href {\doibase 10.1140/epjd/e2004-00130-3} {\bibfield
   {journal} {\bibinfo  {journal} {Eur. Phys. J. D}\ }\textbf {\bibinfo
  {volume} {31}},\ \bibinfo {pages} {365} (\bibinfo {year} {2004})}\BibitemShut
  {NoStop}%
\bibitem [{\citenamefont {Viteau}\ \emph {et~al.}(2008)\citenamefont {Viteau},
  \citenamefont {Chotia}, \citenamefont {Allegrini}, \citenamefont {Bouloufa},
  \citenamefont {Dulieu}, \citenamefont {Comparat},\ and\ \citenamefont
  {Pillet}}]{Viteau2008}%
  \BibitemOpen
  \bibfield  {author} {\bibinfo {author} {\bibfnamefont {M.}~\bibnamefont
  {Viteau}}, \bibinfo {author} {\bibfnamefont {A.}~\bibnamefont {Chotia}},
  \bibinfo {author} {\bibfnamefont {M.}~\bibnamefont {Allegrini}}, \bibinfo
  {author} {\bibfnamefont {N.}~\bibnamefont {Bouloufa}}, \bibinfo {author}
  {\bibfnamefont {O.}~\bibnamefont {Dulieu}}, \bibinfo {author} {\bibfnamefont
  {D.}~\bibnamefont {Comparat}}, \ and\ \bibinfo {author} {\bibfnamefont
  {P.}~\bibnamefont {Pillet}},\ }\bibfield  {title} {\enquote {\bibinfo {title}
  {{Optical pumping and vibrational cooling of molecules.}}}\ }\href {\doibase
  10.1126/science.1159496} {\bibfield  {journal} {\bibinfo  {journal}
  {Science}\ }\textbf {\bibinfo {volume} {321}},\ \bibinfo {pages} {232}
  (\bibinfo {year} {2008})}\BibitemShut {NoStop}%
\bibitem [{\citenamefont {Manai}\ \emph {et~al.}(2012)\citenamefont {Manai},
  \citenamefont {Horchani}, \citenamefont {Lignier}, \citenamefont {Pillet},
  \citenamefont {Comparat}, \citenamefont {Fioretti},\ and\ \citenamefont
  {Allegrini}}]{Manai2012}%
  \BibitemOpen
  \bibfield  {author} {\bibinfo {author} {\bibfnamefont {I.}~\bibnamefont
  {Manai}}, \bibinfo {author} {\bibfnamefont {R.}~\bibnamefont {Horchani}},
  \bibinfo {author} {\bibfnamefont {H.}~\bibnamefont {Lignier}}, \bibinfo
  {author} {\bibfnamefont {P.}~\bibnamefont {Pillet}}, \bibinfo {author}
  {\bibfnamefont {D.}~\bibnamefont {Comparat}}, \bibinfo {author}
  {\bibfnamefont {A.}~\bibnamefont {Fioretti}}, \ and\ \bibinfo {author}
  {\bibfnamefont {M.}~\bibnamefont {Allegrini}},\ }\bibfield  {title} {\enquote
  {\bibinfo {title} {{Rovibrational Cooling of Molecules by Optical
  Pumping}},}\ }\href {\doibase 10.1103/PhysRevLett.109.183001} {\bibfield
  {journal} {\bibinfo  {journal} {Phys. Rev. Lett.}\ }\textbf {\bibinfo
  {volume} {109}},\ \bibinfo {pages} {183001} (\bibinfo {year}
  {2012})}\BibitemShut {NoStop}%
\bibitem [{\citenamefont {Wakim}\ \emph {et~al.}(2012)\citenamefont {Wakim},
  \citenamefont {Zabawa}, \citenamefont {Haruza},\ and\ \citenamefont
  {Bigelow}}]{Wakim2012}%
  \BibitemOpen
  \bibfield  {author} {\bibinfo {author} {\bibfnamefont {A.}~\bibnamefont
  {Wakim}}, \bibinfo {author} {\bibfnamefont {P.}~\bibnamefont {Zabawa}},
  \bibinfo {author} {\bibfnamefont {M.}~\bibnamefont {Haruza}}, \ and\ \bibinfo
  {author} {\bibfnamefont {N.~P.}\ \bibnamefont {Bigelow}},\ }\bibfield
  {title} {\enquote {\bibinfo {title} {{Luminorefrigeration: vibrational
  cooling of NaCs.}}}\ }\href {\doibase 10.1364/OE.20.016083} {\bibfield
  {journal} {\bibinfo  {journal} {Opt. Express}\ }\textbf {\bibinfo {volume}
  {20}},\ \bibinfo {pages} {16083} (\bibinfo {year} {2012})}\BibitemShut
  {NoStop}%
\bibitem [{\citenamefont {Staanum}\ \emph {et~al.}(2010)\citenamefont
  {Staanum}, \citenamefont {H{\o}jbjerre}, \citenamefont {Skyt}, \citenamefont
  {Hansen},\ and\ \citenamefont {Drewsen}}]{Staanum2010}%
  \BibitemOpen
  \bibfield  {author} {\bibinfo {author} {\bibfnamefont {P.~F.}\ \bibnamefont
  {Staanum}}, \bibinfo {author} {\bibfnamefont {K.}~\bibnamefont
  {H{\o}jbjerre}}, \bibinfo {author} {\bibfnamefont {P.~S.}\ \bibnamefont
  {Skyt}}, \bibinfo {author} {\bibfnamefont {A.~K.}\ \bibnamefont {Hansen}}, \
  and\ \bibinfo {author} {\bibfnamefont {M.}~\bibnamefont {Drewsen}},\
  }\bibfield  {title} {\enquote {\bibinfo {title} {{Rotational laser cooling of
  vibrationally and translationally cold molecular ions}},}\ }\href {\doibase
  10.1038/nphys1604} {\bibfield  {journal} {\bibinfo  {journal} {Nat. Phys.}\
  }\textbf {\bibinfo {volume} {6}},\ \bibinfo {pages} {271} (\bibinfo {year}
  {2010})}\BibitemShut {NoStop}%
\bibitem [{\citenamefont {Schneider}\ \emph {et~al.}(2010)\citenamefont
  {Schneider}, \citenamefont {Roth}, \citenamefont {Duncker}, \citenamefont
  {Ernsting},\ and\ \citenamefont {Schiller}}]{Schneider2010}%
  \BibitemOpen
  \bibfield  {author} {\bibinfo {author} {\bibfnamefont {T.}~\bibnamefont
  {Schneider}}, \bibinfo {author} {\bibfnamefont {B.}~\bibnamefont {Roth}},
  \bibinfo {author} {\bibfnamefont {H.}~\bibnamefont {Duncker}}, \bibinfo
  {author} {\bibfnamefont {I.}~\bibnamefont {Ernsting}}, \ and\ \bibinfo
  {author} {\bibfnamefont {S.}~\bibnamefont {Schiller}},\ }\bibfield  {title}
  {\enquote {\bibinfo {title} {{All-optical preparation of molecular ions in
  the rovibrational ground state}},}\ }\href {\doibase 10.1038/nphys1605}
  {\bibfield  {journal} {\bibinfo  {journal} {Nat. Phys.}\ }\textbf {\bibinfo
  {volume} {6}},\ \bibinfo {pages} {275} (\bibinfo {year} {2010})}\BibitemShut
  {NoStop}%
\bibitem [{\citenamefont {Rellergert}\ \emph {et~al.}(2013)\citenamefont
  {Rellergert}, \citenamefont {Sullivan}, \citenamefont {Schowalter},
  \citenamefont {Kotochigova}, \citenamefont {Chen},\ and\ \citenamefont
  {Hudson}}]{Rellergert2013}%
  \BibitemOpen
  \bibfield  {author} {\bibinfo {author} {\bibfnamefont {W.~G.}\ \bibnamefont
  {Rellergert}}, \bibinfo {author} {\bibfnamefont {S.~T.}\ \bibnamefont
  {Sullivan}}, \bibinfo {author} {\bibfnamefont {S.~J.}\ \bibnamefont
  {Schowalter}}, \bibinfo {author} {\bibfnamefont {S.}~\bibnamefont
  {Kotochigova}}, \bibinfo {author} {\bibfnamefont {K.}~\bibnamefont {Chen}}, \
  and\ \bibinfo {author} {\bibfnamefont {E.~R.}\ \bibnamefont {Hudson}},\
  }\bibfield  {title} {\enquote {\bibinfo {title} {{Evidence for sympathetic
  vibrational cooling of translationally cold molecules.}}}\ }\href {\doibase
  10.1038/nature11937} {\bibfield  {journal} {\bibinfo  {journal} {Nature}\
  }\textbf {\bibinfo {volume} {495}},\ \bibinfo {pages} {490} (\bibinfo {year}
  {2013})}\BibitemShut {NoStop}%
\bibitem [{\citenamefont {Hansen}\ \emph {et~al.}(2014)\citenamefont {Hansen},
  \citenamefont {Versolato}, \citenamefont {K{\l}osowski}, \citenamefont
  {Kristensen}, \citenamefont {Gingell}, \citenamefont {Schwarz}, \citenamefont
  {Windberger}, \citenamefont {Ullrich}, \citenamefont {Crespo
  L\'{o}pez-Urrutia},\ and\ \citenamefont {Drewsen}}]{Hansen2014}%
  \BibitemOpen
  \bibfield  {author} {\bibinfo {author} {\bibfnamefont {A.~K.}\ \bibnamefont
  {Hansen}}, \bibinfo {author} {\bibfnamefont {O.~O.}\ \bibnamefont
  {Versolato}}, \bibinfo {author} {\bibfnamefont {L.}~\bibnamefont
  {K{\l}osowski}}, \bibinfo {author} {\bibfnamefont {S.~B.}\ \bibnamefont
  {Kristensen}}, \bibinfo {author} {\bibfnamefont {A.}~\bibnamefont {Gingell}},
  \bibinfo {author} {\bibfnamefont {M.}~\bibnamefont {Schwarz}}, \bibinfo
  {author} {\bibfnamefont {A.}~\bibnamefont {Windberger}}, \bibinfo {author}
  {\bibfnamefont {J.}~\bibnamefont {Ullrich}}, \bibinfo {author} {\bibfnamefont
  {J.~R.}\ \bibnamefont {Crespo L\'{o}pez-Urrutia}}, \ and\ \bibinfo {author}
  {\bibfnamefont {M.}~\bibnamefont {Drewsen}},\ }\bibfield  {title} {\enquote
  {\bibinfo {title} {{Efficient rotational cooling of Coulomb-crystallized
  molecular ions by a helium buffer gas.}}}\ }\href {\doibase
  10.1038/nature12996} {\bibfield  {journal} {\bibinfo  {journal} {Nature}\
  }\textbf {\bibinfo {volume} {508}},\ \bibinfo {pages} {76} (\bibinfo {year}
  {2014})}\BibitemShut {NoStop}%
\bibitem [{\citenamefont {Lien}\ \emph {et~al.}(2014)\citenamefont {Lien},
  \citenamefont {Seck}, \citenamefont {Lin}, \citenamefont {Nguyen},
  \citenamefont {Tabor},\ and\ \citenamefont {Odom}}]{Lien2014}%
  \BibitemOpen
  \bibfield  {author} {\bibinfo {author} {\bibfnamefont {C.-Y.}\ \bibnamefont
  {Lien}}, \bibinfo {author} {\bibfnamefont {C.~M.}\ \bibnamefont {Seck}},
  \bibinfo {author} {\bibfnamefont {Y.-W.}\ \bibnamefont {Lin}}, \bibinfo
  {author} {\bibfnamefont {J.~H.~V.}\ \bibnamefont {Nguyen}}, \bibinfo {author}
  {\bibfnamefont {D.~A.}\ \bibnamefont {Tabor}}, \ and\ \bibinfo {author}
  {\bibfnamefont {B.~C.}\ \bibnamefont {Odom}},\ }\bibfield  {title} {\enquote
  {\bibinfo {title} {{Broadband optical cooling of molecular rotors from room
  temperature to the ground state.}}}\ }\href {\doibase 10.1038/ncomms5783}
  {\bibfield  {journal} {\bibinfo  {journal} {Nat. Commun.}\ }\textbf {\bibinfo
  {volume} {5}},\ \bibinfo {pages} {4783} (\bibinfo {year} {2014})}\BibitemShut
  {NoStop}%
\bibitem [{\citenamefont {Zeppenfeld}\ \emph {et~al.}(2009)\citenamefont
  {Zeppenfeld}, \citenamefont {Motsch}, \citenamefont {Pinkse},\ and\
  \citenamefont {Rempe}}]{Zeppenfeld2009}%
  \BibitemOpen
  \bibfield  {author} {\bibinfo {author} {\bibfnamefont {M.}~\bibnamefont
  {Zeppenfeld}}, \bibinfo {author} {\bibfnamefont {M.}~\bibnamefont {Motsch}},
  \bibinfo {author} {\bibfnamefont {P.~W.~H.}\ \bibnamefont {Pinkse}}, \ and\
  \bibinfo {author} {\bibfnamefont {G.}~\bibnamefont {Rempe}},\ }\bibfield
  {title} {\enquote {\bibinfo {title} {{Optoelectrical cooling of polar
  molecules}},}\ }\href {\doibase 10.1103/PhysRevA.80.041401} {\bibfield
  {journal} {\bibinfo  {journal} {Phys. Rev. A}\ }\textbf {\bibinfo {volume}
  {80}},\ \bibinfo {pages} {041401} (\bibinfo {year} {2009})}\BibitemShut
  {NoStop}%
\bibitem [{\citenamefont {Englert}\ \emph {et~al.}(2011)\citenamefont
  {Englert}, \citenamefont {Mielenz}, \citenamefont {Sommer}, \citenamefont
  {Bayerl}, \citenamefont {Motsch}, \citenamefont {Pinkse}, \citenamefont
  {Rempe},\ and\ \citenamefont {Zeppenfeld}}]{Englert2011}%
  \BibitemOpen
  \bibfield  {author} {\bibinfo {author} {\bibfnamefont {B.~G.~U.}\
  \bibnamefont {Englert}}, \bibinfo {author} {\bibfnamefont {M.}~\bibnamefont
  {Mielenz}}, \bibinfo {author} {\bibfnamefont {C.}~\bibnamefont {Sommer}},
  \bibinfo {author} {\bibfnamefont {J.}~\bibnamefont {Bayerl}}, \bibinfo
  {author} {\bibfnamefont {M.}~\bibnamefont {Motsch}}, \bibinfo {author}
  {\bibfnamefont {P.~W.~H.}\ \bibnamefont {Pinkse}}, \bibinfo {author}
  {\bibfnamefont {G.}~\bibnamefont {Rempe}}, \ and\ \bibinfo {author}
  {\bibfnamefont {M.}~\bibnamefont {Zeppenfeld}},\ }\bibfield  {title}
  {\enquote {\bibinfo {title} {{Storage and Adiabatic Cooling of Polar
  Molecules in a Microstructured Trap}},}\ }\href {\doibase
  10.1103/PhysRevLett.107.263003} {\bibfield  {journal} {\bibinfo  {journal}
  {Phys. Rev. Lett.}\ }\textbf {\bibinfo {volume} {107}},\ \bibinfo {pages}
  {263003} (\bibinfo {year} {2011})}\BibitemShut {NoStop}%
\bibitem [{\citenamefont {Zeppenfeld}(2013)}]{Zeppenfeld2013}%
  \BibitemOpen
  \bibfield  {author} {\bibinfo {author} {\bibfnamefont {M.}~\bibnamefont
  {Zeppenfeld}},\ }{\enquote  {\bibinfo {title} {{Electric Trapping and Cooling of
  Polyatomic Molecules}},}\ }\href@noop {} {Ph.D. thesis},\ \bibinfo  {school}
  {Technical University of Munich} (\bibinfo {year} {2013})\BibitemShut
  {NoStop}%
\bibitem [{\citenamefont {Gl\"{o}ckner}\ \emph {et~al.}(2015)\citenamefont
  {Gl\"{o}ckner}, \citenamefont {Prehn}, \citenamefont {Rempe},\ and\
  \citenamefont {Zeppenfeld}}]{Gloeckner2015}%
  \BibitemOpen
  \bibfield  {author} {\bibinfo {author} {\bibfnamefont {R.}~\bibnamefont
  {Gl\"{o}ckner}}, \bibinfo {author} {\bibfnamefont {A.}~\bibnamefont {Prehn}},
  \bibinfo {author} {\bibfnamefont {G.}~\bibnamefont {Rempe}}, \ and\ \bibinfo
  {author} {\bibfnamefont {M.}~\bibnamefont {Zeppenfeld}},\ }\bibfield  {title}
  {\enquote {\bibinfo {title} {{Rotational state detection of electrically
  trapped polyatomic molecules}},}\ }\href
  {http://stacks.iop.org/1367-2630/17/i=5/a=055022} {\bibfield  {journal}
  {\bibinfo  {journal} {New J. Phys.}\ }\textbf {\bibinfo {volume} {17}},\
  \bibinfo {pages} {055022} (\bibinfo {year} {2015})}\BibitemShut {NoStop}%
\bibitem [{NoteSup()}]{NoteSup}%
  \BibitemOpen
  \bibinfo {note} {See Supplemental Material}\BibitemShut {NoStop}%
\bibitem [{\citenamefont {Watson}(1971)}]{Watson1971}%
  \BibitemOpen
  \bibfield  {author} {\bibinfo {author} {\bibfnamefont {J.~K.}\ \bibnamefont
  {Watson}},\ }\bibfield  {title} {\enquote {\bibinfo {title} {{Forbidden
  rotational spectra of polyatomic molecules}},}\ }\href {\doibase
  10.1016/0022-2852(71)90255-4} {\bibfield  {journal} {\bibinfo  {journal} {J.
  Mol. Spectrosc.}\ }\textbf {\bibinfo {volume} {40}},\ \bibinfo {pages} {536}
  (\bibinfo {year} {1971})}\BibitemShut {NoStop}%
\end{thebibliography}
\end{document}